\documentclass[reprint,aip,jap,numerical,superscriptaddress]{revtex4-1}

\usepackage{amsmath}
\usepackage{amssymb}
\usepackage[pdftex]{graphicx}
\usepackage{dcolumn}
\usepackage{bm}

\begin{document}
\title{Long-lived nanosecond spin coherence in high-mobility 2DEGs \\ confined in double and triple quantum wells}

\author{S. Ullah}
\author{G. M. Gusev}
\affiliation {Instituto de F\'{i}sica, Universidade de S\~{a}o Paulo, Caixa Postal 66318 - CEP 05315-970, S\~{a}o Paulo, SP, Brazil}
\author{A. K. Bakarov}
\affiliation{Institute of Semiconductor Physics and Novosibirsk State University, Novosibirsk 630090, Russia}
\author{F. G. G. Hernandez}
 \email[Corresponding author.\\ Electronic address: ]{felixggh@if.usp.br}
\affiliation {Instituto de F\'{i}sica, Universidade de S\~{a}o Paulo, Caixa Postal 66318 - CEP 05315-970, S\~{a}o Paulo, SP, Brazil}
\date{\today}

\begin{abstract}
We investigated the spin coherence of high-mobility two-dimensional electron gases confined in multilayer GaAs quantum wells. The dynamics of the spin polarization was optically studied using pump-probe techniques: time-resolved Kerr rotation and resonant spin amplification. For double and triple quantum wells doped beyond the metal-to-insulator transition, the spin-orbit interaction was tailored by the sample parameters of structural symmetry (Rashba constant), width and electron density (Dresselhaus linear and cubic constants) which allows us to attain long dephasing times in the nanoseconds range. The determination of the scales: transport scattering time, single-electron scattering time, electron-electron scattering time, and spin polarization decay time further supports the possibility of using n-doped multilayer systems for developing spintronic devices.
\end{abstract}

\maketitle

\section{Introduction}

Long-lived spin coherence time ($T_{2}^{*}$) for ensembles is a milestone for practical applications of spintronic devices.\cite{wu} The tunability of $T_{2}^{*}$ have been widely studied in semiconductor quantum wells (QWs) with a large variety of optical techniques developed for the study of spin polarization dynamics and spin relaxation mechanisms.\cite{5,6,7,8} In n-type samples, for example, it was observed that the doping level has a major role to attain long coherence time or to limit it with $T_{2}^{*}$ changing from tens of picoseconds up to nanoseconds.\cite{13,22,28,30} The turning point, where $T_{2}^{*}$ decreases with an increase of the electron concentration, was found at the metal-to-insulator transition for bulk\cite{dzhioev02,romer} ($2\times10^{16}$ cm$^{-3}$) and GaAs QWs\cite{sandhu} ($5\times10^{10}$ cm$^{-2}$). Beyond this point, the Dyakonov-Perel (DP) spin relaxation mechanism is dominant and controlled by electron-electron collisions.\cite{leyland}

The DP mechanism defines that the decay time of the spin polarization $t_z$ (along the QW out-of-plane direction) is limited by the spin-orbit interaction which give us a path to control spin coherence. It can be calculated according to: $t_z^{-1}=8D_sm^2\hbar^{-4}[\alpha^{2}+(\beta_1-\beta_3)^2+\beta_3^2)]$, where $D_s$ is the spin diffusion constant, $\alpha$ is the Rashba coefficient due to structural inversion asymmetry, and $\beta_1$ and $\beta_3$ are the linear and cubic Dresselhaus constants due to bulk inversion asymmetry.\cite{walser,kainz}

Recently, the authors demonstrated that multilayer QWs are exceptional platforms for the investigation of current-induced spin polarization effects.\cite{31,3} While such complex systems also offer new possibilities for applications, for example in the production of spin blockers\cite{souma} and filters,\cite{23} the study of long-lived spin coherence in double (DQW) and triple quantum wells (TQW) is still required. Here, we report on the coherent spin dynamics in multilayer quantum wells using time-resolved Kerr rotation (TRKR) and resonant spin amplification (RSA). The sample structure allowed us to tailor the spin-orbit constants by the well width, symmetry and subband concentration parameters. Remarkably, it results in coherence times in the nanoseconds range even for DQW and TQW samples with individual subband density beyond the metal-to-insulator transition.

\section{Materials and Experiment}

We investigated two different samples grown in the [001] direction, one double and one triple quantum well, both containing a dense two-dimensional electron gas (2DEG) with equal total density. For both samples, the barriers were made of short-period AlAs/GaAs superlattices (SPSL) in order to shield the doping ionized impurities and efficiently enhance the mobility.\cite{bakarov} The density of the Si delta-doping was $2.2\times10^{12}$cm$^{-2}$ symmetrically separated from the QW by 7 periods of the SPSL with 4 AlAs monolayers and 8 GaAs monolayers per period. The DQW sample consists of a wide doped GaAs well with w $=45$ nm, total electron density $n_{t}=9.2\times10^{11}$cm$^{-2}$ and mobility $\mu=1.9\times10^{6}$cm$^2$/Vs at low temperature. The electronic system has a DQW configuration with symmetric and antisymmetric wave functions for the two lowest subbands with subband separation $\Delta_{12}=1.4$ meV and approximately equal subband density $n_{s}$.\cite{wiedmannDQW} Fig. 1(a) shows the calculated DQW band structure and the charge density for both subbands. 

The second sample is a symmetrically doped GaAs TQW with 2 nm-thick $Al_{0.3}Ga_{0.7}As$ barriers, $n_{t}=9\times10^{11}$cm$^{-2}$ and $\mu=5\times10^{5}$cm$^2$/Vs measured at low temperature. The central well width is 22 nm and both side wells have equal width of 10 nm. The central well has a larger width in order to be populated because the electron density tends to concentrate mostly in the side wells as a result of electron repulsion and confinement. The estimated density in the central well is 35\% smaller than in the side wells. The coupling strength between the quantum wells is characterized by the separation energies $\Delta_{ij}$ of the three occupied subbands ($i,j=1,2,3$) given by $\Delta_{12}=1.0$ meV, $\Delta_{23}=2.4$ meV, $\Delta_{13}=3.4$ meV.\cite{wiedmannTQW}

TRKR and RSA were used to probe the coherent spin dynamics in the electron gas. For optical excitation, we used a mode-locked Ti-sapphire laser with pulse duration of 100 fs and repetition rate of $f_{rep}=$ 76 MHz corresponding to a repetition period ($t_{rep}$) of 13.2 ns. The time delay $\Delta t$ between pump and probe pulses was varied by a mechanical delay line. The pump beam was circularly polarized by means of a photo-elastic modulator operated at a frequency of 50 kHz. The rotation of the probe polarization was recorded as function of $\Delta t$ and detected with a balanced bridge using coupled photodiodes. The laser wavelength was tuned looking for the TRKR energy dependence in each sample. The samples were immersed in the variable temperature insert of a split-coil superconductor magnet in the Voigt geometry.

\section{Results and Discussion}

The time evolution of the spin dynamics for the DQW is displayed in Fig. 1(b) up to 2 T with pump/probe power of 1 mW/300 $\mu$W. The TRKR oscillations are associated with the precession of coherently excited electron spins about an in-plane magnetic field. To obtain the spin coherence time, the evolution of the Kerr rotation angle can be described by an exponentially damped harmonic:
\begin{equation}
\theta_{K}(\Delta t) = A \exp(-\Delta t/T_{2}^{*})\cos(\omega_{L}\Delta t + \phi)
\end{equation}
where $A$ is the initial spin polarization build-up by the pump, $\phi$ is the oscillation phase, and $\omega_{L}=g\mu_{B}$B/$\hbar$ is the Larmor frequency with magnetic field B, electron g-factor (absolute value) $g$, Bohr magneton $\mu_{B}$, and reduced Planck's constant $\hbar$. The magnetic field dependence of $\omega_{L}$ and T$_{2}^{*}$ are shown in Fig. 1(c) and (d). Solid lines are fits to the data. One can clearly see that the spin precession frequency increases with B following the linear dependence of the Larmor frequency on the applied field. The value of the fitted g-factor is 0.453 which is close to absolute value for bulk GaAs and similar to the value measured for a quase-two-dimensional system in a single barrier heterostructure with two-subbands occupied.\cite{zhangEPL}

\begin{figure}[h!]
\includegraphics[width=1\columnwidth]{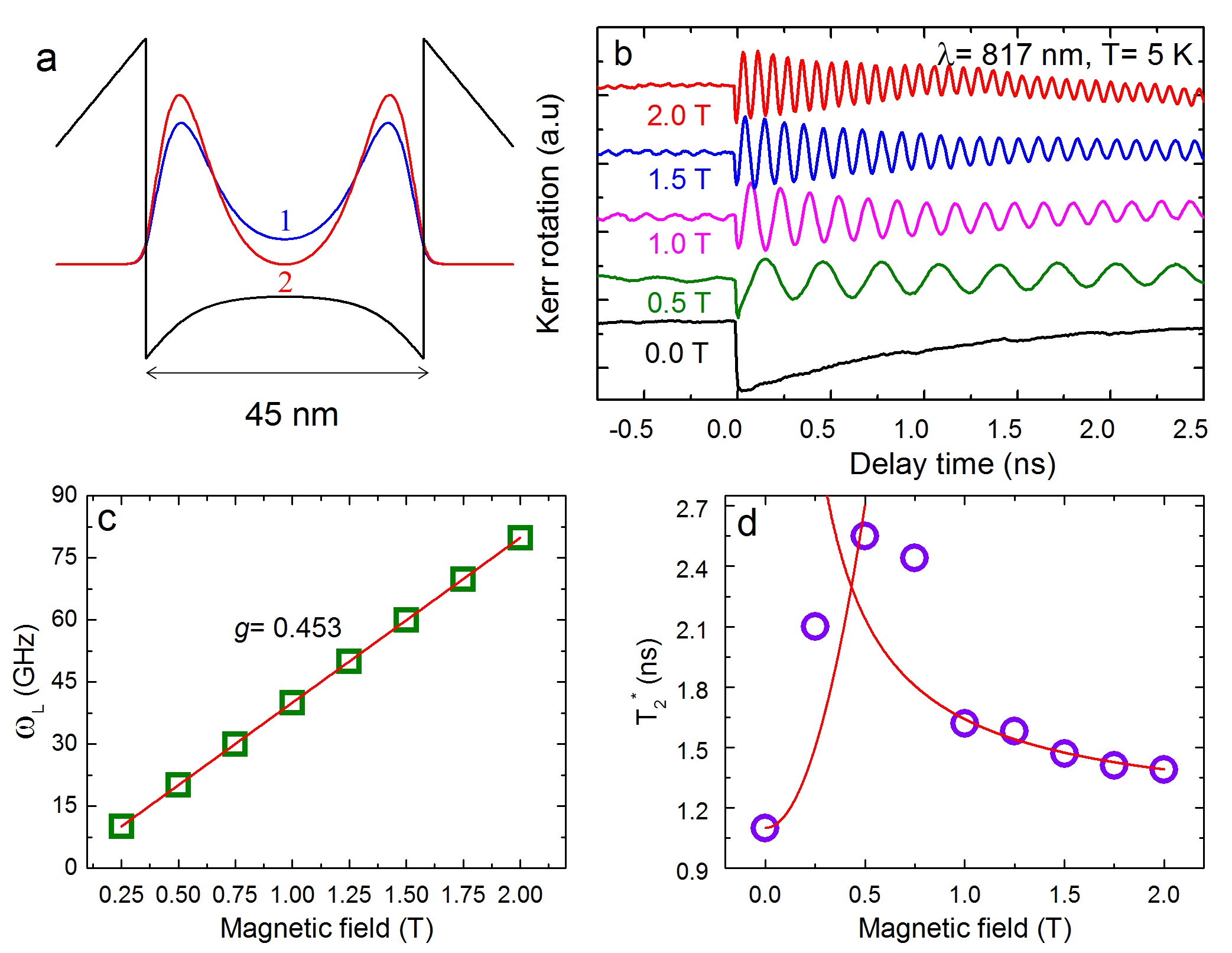}
\caption{(a) DQW band structure and charge density for the first and second subbands. (b) KR as function of the pump-probe delay for different magnetic fields. (c) Larmor frequency $\omega_L$ and (d) T$_{2}^{*}$ fitted as function of B.}
\end{figure}

According to the Dyakonov-Perel mechanism, the observed exponential decay at B $=$ 0 corresponds to the strong scattering regime. In the opposite case, where the spin precess more than a revolution before being scattered, an oscillatory behavior would be expected.\cite{leyland,brand} The measured value for the decay time of the spin polarization along the z-direction (out-of-plane) is 1.1 ns at zero external field. For our symmetric, wide and dense quantum well; we estimate $\alpha\simeq0$, $\beta_1=-\gamma(\pi/w)^2=0.49\times10^{-13}$eVm, and $\beta_3=-\frac{1}{2}\gamma\pi n_s=0.70\times10^{-13}$eVm for the first subband using $\gamma=-$10 eV$\AA^3$.\cite{walserprb} The charge diffusion constant can be estimated, using the effective mass $m$ and the electron's charge $e$, from the Fermi velocity $v_F=\hbar\sqrt{2\pi n_s}/m$ and the transport scattering time $\tau=\mu m/e=70$ ps by $D_c=v_F^2\tau/2=$ 3 m$^2$/s. The diffusion constant for spins is approximately two orders of magnitude smaller than for charge.\cite{walser} Scaling $D_s=100$ to 300 cm$^2$/s, we obtain $t_z\sim[8D_sm^2\hbar^{-4}\beta_3^2]^{-1}=$ 1.1 to 3.3 ns. The data at B $=$ 0 thus agrees with a DP mechanism where the spin dynamics is dominated by the cubic Dresselhaus term. The cancellation of $\alpha\simeq0$ and $\beta_1-\beta_3\simeq0$ due to the sample parameters shows a practical path for long-lived spin coherence in highly doped QWs.

Increasing the magnetic field up to 0.5 T, we found a systematic increase of T$_{2}^{*}$. In this situation, the cyclotron motion acts as momentum scattering and leads to a less efficient spin relaxation in agreement with the DP model.\cite{griesbeck} It is important to note that the in-plane magnetic field was applied using Voigt configuration and the  the cyclotron motion is perpendicular to the QW plane. The increase follows a quadratic dependence\cite{41} with $T_{2}^{*}(B)/T_{2}^{*}(0)=1+(\omega_{c}\tau_p^*)^2$ where $\omega_{c}$ is the cyclotron frequency and $\tau_p^*$ is the single-electron momentum scattering time. We found $\tau_p^*=$ 0.92 ps in agreement with the magnitude of the quantum lifetime measured by transport from the Dingle factor of the magneto-intersubband oscillations on the same sample.\cite{wiedmann2010} The value for $\tau_p^*$ is also in agreement with the determination of approximately 0.5 ps for QWs of shorter width.\cite{walserprb} One of the reasons for the large difference between $\tau$ and $\tau_p^*$ is the insensibility of the first to electron-electron scattering. The ratio of $\tau/\tau_p^*\simeq100$ implies that the dominant scattering from impurities is due to remote instead of background impurities.\cite{macleod} If we consider that $1/\tau_p^*=1/\tau+1/\tau_{ee}$, we get a time scale of $\tau_{ee}=\tau_p^*$ which demonstrates that electron-electron collisions dominate the microscopic scattering mechanisms as expected.\cite{leyland}

Additionally, a further increase of the magnetic field leads to a strong decay due to the spread of the g-factor within the measured ensemble.\cite{greilichprl2006,zhukov2007} The size of the inhomogeneity $\Delta g$ can be inferred by fitting the data according to $1/T_{2}^{*}(B)=1/T_{2}^{*}(0)+\Delta g\mu_BB/\sqrt{2}\hbar$ as shown in Fig. 1(d). From the 1/B dependence,\cite{22,28} we obtain $\Delta g=0.002$ or 0.44\% and $T_{2}^{*}(0)=$ 2 ns.

The optical power influence on the spin dynamics for the DQW sample is shown in Fig. 2(a) at 1 T. Only at low pump power, we observed negative delay oscillations of considerably large amplitude. To find electron spin polarization before the pump pulse arrival indicates that the spin polarization persists from the previous pump pulse, which took place $t_{rep}=13.2$ ns before. The excitation power dependence of T$_{2}^{*}$ was plotted in Fig. 2(b) yielding an exponential decay. For single QW structures, the decrease of the coherence time at high pump density was associated with the electrons delocalization caused by their heating due to the interaction with the photogenerated carriers.\cite{zhukov2007} A similar decrease was also attributed to an increased efficiency of the Bir-Aronov-Pikus mechanism induced by the larger hole photogenerated density in GaAs QWs.\cite{37} However, it is unlikely to be relevant in our dense 2DEG where the photogenerated hole loses its spin and energy quickly and fast recombines with an electron from the 2DEG. Nevertheless, being a key parameter for spin devices, we note that T$_{2}^{*}$ remains near the nanoseconds range when the power is raised by almost one order of magnitude. At higher excitation power, an additional short-lived component in the signal becomes more significant. In systems where T$_{2}^{*}$ is comparable or longer than the laser repetition period, one can use the RSA technique to extract the spin dephasing time by scanning the magnetic field at a fixed pump-probe delay.\cite{22} We note that the 2DEG dynamics is associated with the long lasting oscillations, rather than with excitons or photo-excited electrons.\cite{zhukov2007}

\begin{figure}[h!]
\includegraphics[width=1\columnwidth]{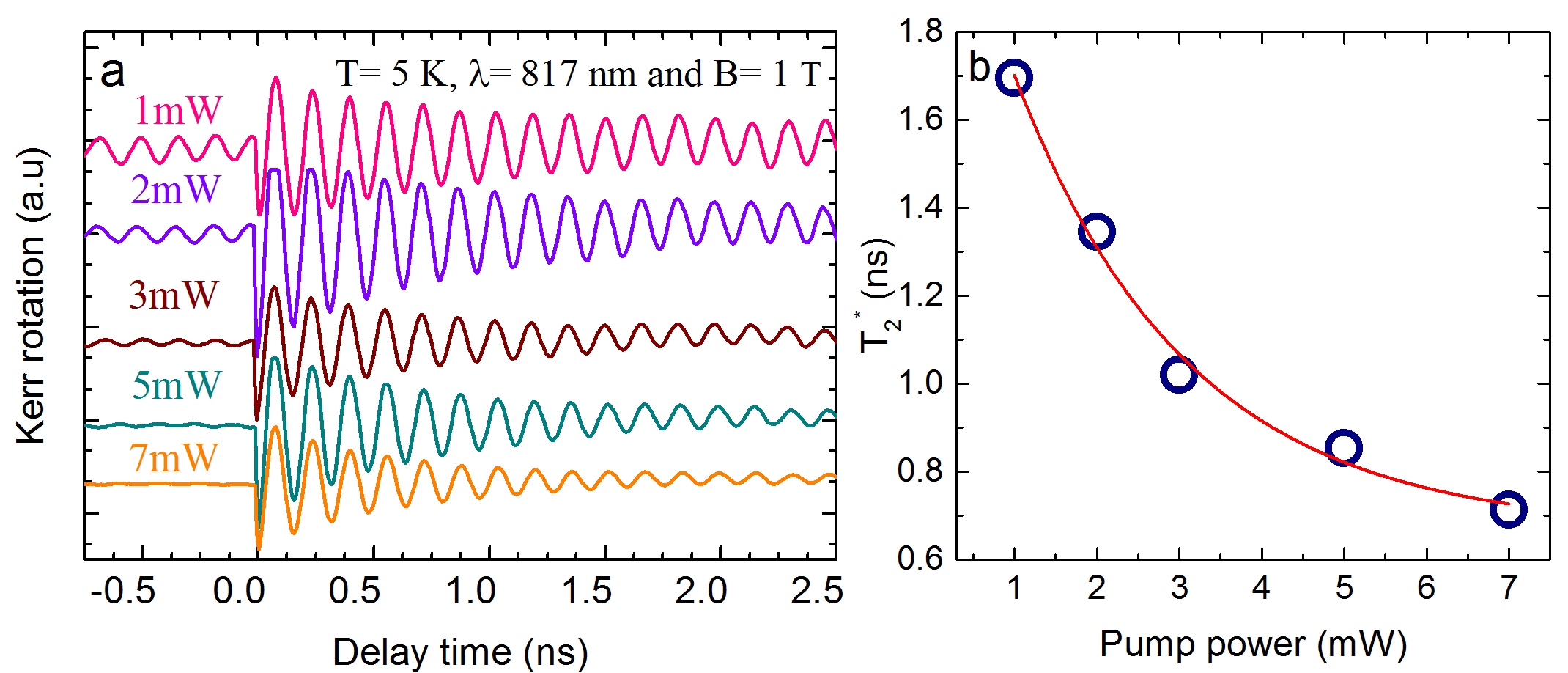}
\caption{(a) TRKR of the DQW as function of pump power and (b) the corresponding T$_{2}^{*}$.}
\end{figure}

\begin{figure}[h!]
\includegraphics[width=1\columnwidth]{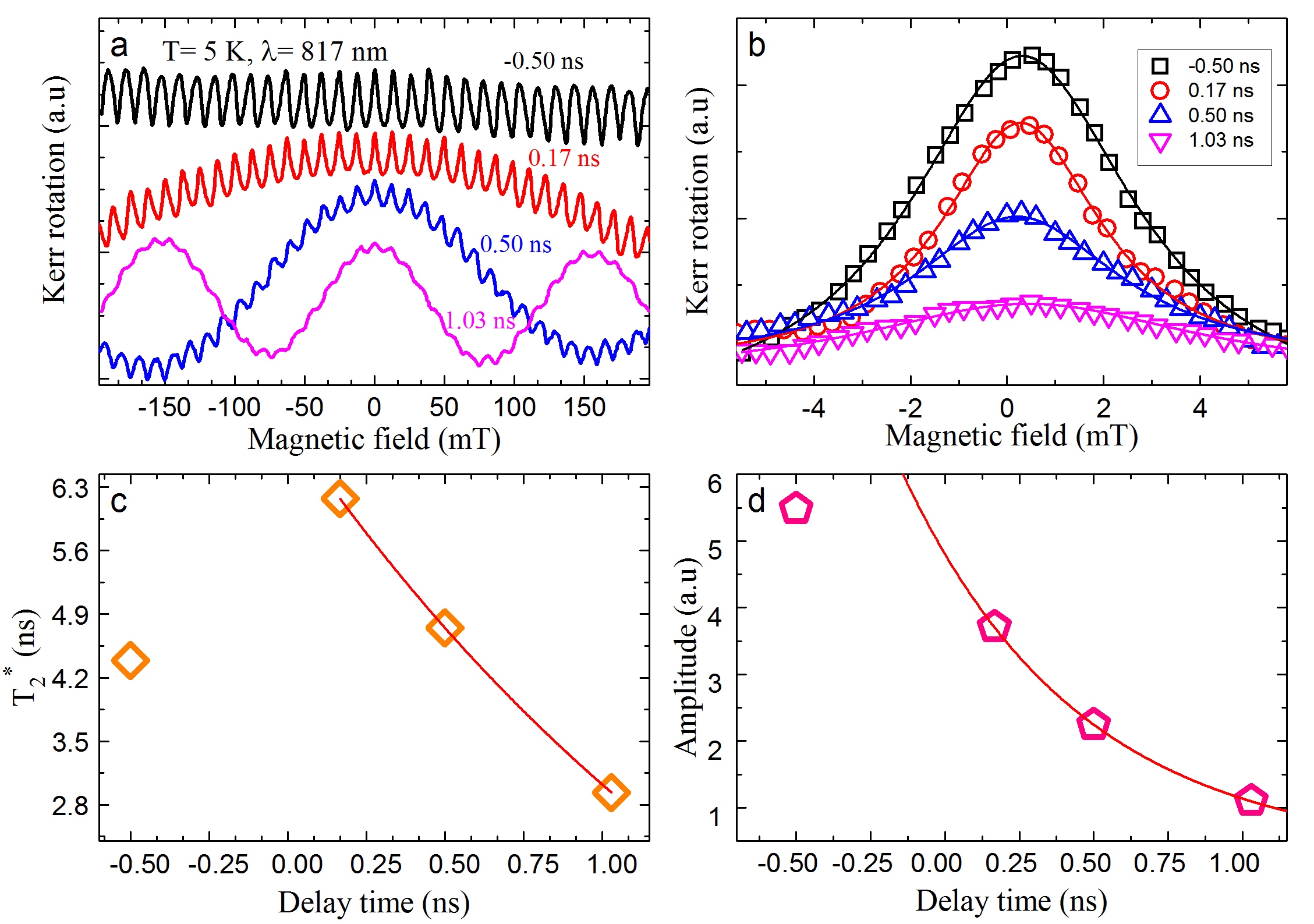}
\caption{(a) RSA scans of the DQW system obtained for different time delays. (b) Lorentzian fit of the zero-field resonance peak. (c) T$_{2}^{*}$ and (d) Amplitude dependence on $\Delta t$ from (b).}
\end{figure}

Fig. 3(a) displays the RSA signals measured for different $\Delta t$ with pump/probe power of 1 mW/300 $\mu$W. We observed a series of sharp resonance peaks as a function of B corresponding to the electron spin precession frequencies which are commensurable with the pump pulse repetition period obeying the periodic condition: $\Delta B = (h f_{rep})/(g \mu_B)$.\cite{22} As function of the magnetic field, the RSA peaks amplitude decreases as a result of the g-factor variation within the measured ensemble as noted above. The RSA resonances are modulated by a slow oscillation that depends on $f_{d}=1/\Delta t$ according to the same periodic condition. We will focus on the zero field resonance. T$_{2}^{*}$ can be directly evaluated from the width of the zero-th resonance using a Hanle (Lorentzian) model:\cite{3,22} 
\begin{equation}
\theta_{K}=A/[(\omega_{L}T_{2}^{*})^{2}+1]
\end{equation} 
with half-width $B_{1/2}=\hbar/(g\mu_{B}T_{2}^{*})$. The fitting result is displayed in Fig. 3(b) for negative and positive delays. The extracted values for T$_{2}^{*}$ and for the amplitude are shown in Fig. 3(c) and (d) as function of $\Delta t$. For positive delays, both quantities display an exponential decay (solid line). Increasing the pump-probe delay cause the broadening of RSA peaks according to a shorter spin dephasing time. However, the system coherence is recovered just before pump arrival for the long-lived spin component in the system dynamics.\cite{yugova2012} The RSA amplitude measured at negative delay was T$_{2}^{*}=$ 4.4 ns.

\begin{figure}[h!]
\includegraphics[width=1\columnwidth]{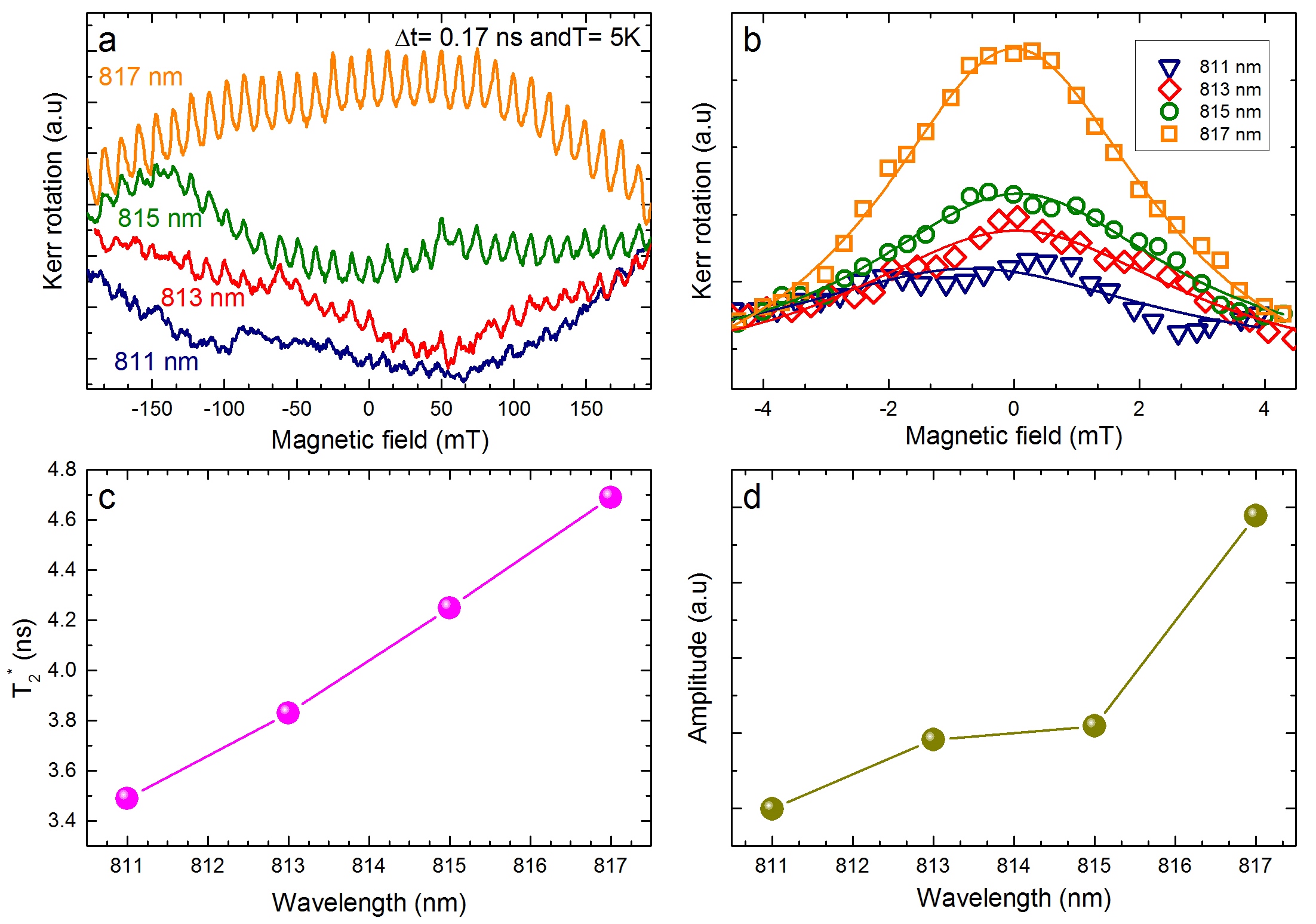}
\caption{(a) RSA scans of the DQW sample measured for different pump-probe wavelengths. (b) Fitting of the zero-field resonance. (c) Spin coherence time T$_{2}^{*}$ and (d) Amplitude extracted from (b).}
\end{figure}

Concerning the subband dependence, the spin relaxation time was calculated to be identical in an electron system with two occupied subbands, although the higher subband may have a much larger inhomogeneous broadening, due to strong intersubband Coulomb scattering.\cite{weng,zhangEPL} In our samples, we studied the pump/probe wavelength dependence as reported in TRKR\cite{zhangEPL} and photoluminescence\cite{pussep2,pussep3} studies on similar multilayer systems. 

Figure 4(a) displays the RSA scans of the DQW sample for different pump-probe wavelengths at fixed delay. Panel 4(b) shows a comparison between the zero-field resonances where the solid line is a Hanle fit to the data as described above. T$_{2}^{*}$ and the amplitude obtained from (b) increase with the pump-probe wavelength as shown in Figures 4(c) and 4(d). Increasing the pump-probe energy about 3 meV ($\simeq2\Delta_{12}$), from 817 nm to 815 nm, leads to a T$_{2}^{*}$ decrease of less than 10\% in Figure 4(d). In comparison, Figure 1 shows negative delay oscillations in the same wavelength range.\cite{SM} This small change could be associated with the relative similitude between the charge density distribution for both subbands. On the other side, fast intersubband scattering may be hiding differences expected in the spin-orbit interaction for the second subband.\cite{egues}

\begin{figure}[h!]
\includegraphics[width=1\columnwidth]{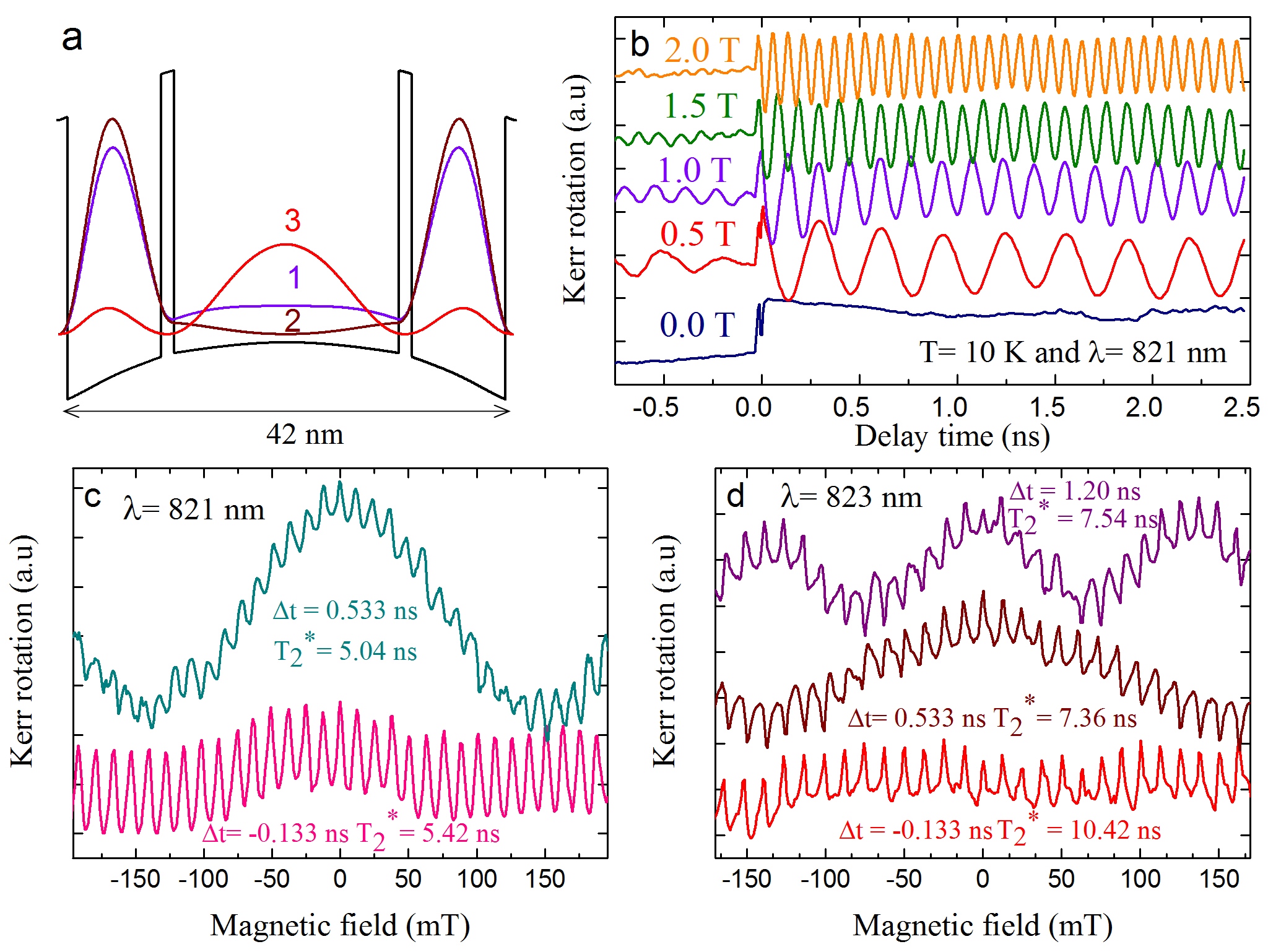}
\caption{(a) Band diagram and charge density for the TQW sample. (b) TRKR measured as function of the magnetic field. RSA scans of the TQW sample measured for different pump-probe delays with the corresponding extracted spin dephasing time at (c) 821 nm and (d) 823 nm.}
\end{figure}

Finally, we focus on the results for the TQW sample. Fig. 5(a) shows the calculated band diagram and charge density for three occupied subbands. The TRKR scans measured as function of the magnetic field yield g $=$ 0.452. Due to the long spin coherence comparable with the laser repetition period, there is almost no decay over the measured time window (2.5 ns). In analogy to the DQW sample, we used the constructive interference of the coherence oscillations from successive pulses to extract the spin coherence time by the RSA technique. Fig. 5(c) and (d) show the magnetic field scans of the KR amplitude performed at different pump/probe separation for 821 and 823 nm, respectively. From the Lorentzian fit of the zero-field peak, the spin dephasing for the TQW sample was obtained revealing the longest T$_{2}^{*}=$ 10.42 ns at negative delay as for the DQW. In this case, the same energy increase ($\sim$3 meV $\simeq\Delta_{13}$), leads to strong T$_{2}^{*}$ decrease of almost 50\%/30\% at negative/positive delay. We note that, contrary to the DQW case, the third subband for the TQW have opposite charge distribution if compared with the lower subbands. While the third subband has the charge density more localized in the central well, the electrons in the first and second subbands are distributed in the side wells. 

\section{Conclusions}

In conclusion, we have studied the spin dynamics of a two-dimensional electron gas in multilayer QWs by TRKR and RSA. The dependence of spin dephasing time on the experimental parameters: magnetic field, pump power, and pump-probe delay was demonstrated. In the DQW sample, T$_{2}^{*}$ extends to 4.4 ns. Additionally, for the TQW sample, T$_{2}^{*}$ exceeding 10 ns was observed. The results found are among the longest T$_{2}^{*}$ reported for samples of similar doping level \cite{sandhu,41} and comparable with nominally undoped narrow GaAs QWs \cite{dzhioev} and low density 2DEGs in CdTe QWs \cite{zhukov2007}. The measured long spin dephasing time was tailored by the control of the QW width, symmetry and electron density. The spin dynamics is dominated through the cubic Dresselhaus interaction by the DP mechanism. All the relevant time scales were determined indicating the importance of each scattering mechanism in the spin dynamics. We demonstrate that the wave function engineering in multilayer QWs may provide practical paths to control the dynamics in spintronic devices.

\section{Acknowledgments}

F.G.G.H. acknowledges financial support from Grants No. 2009/15007-5, No. 2013/03450-7, and No. 2014/25981-7 of the S\~{a}o Paulo Research Foundation (FAPESP). S.U. acknowledges TWAS/CNPq for financial support.

\bibliographystyle{apsrev}
\bibliography{JAP_FGGH_final}
\end{document}